\documentstyle[12pt]{article}

\def\Journal#1#2#3#4{{#1} {\bf #2}, #3 (#4)}

\def\NPB{{\em Nucl. Phys.} B}
\def\PLB{{\em Phys. Lett.}  B}

\def\PRD{{\em Phys. Rev.} D}
\def\ZPC{{\em Z. Phys.} C}

\def\be{\begin{equation}}
\def\ee{\end{equation}}
\def\bea{\begin{eqnarray}}
\def\eea{\end{eqnarray}}

\begin{document}
\hfill{NCKU-HEP-97-07}
\vskip 0.5cm
\centerline{\large\bf SOME ASPECTS OF THE BFKL EVOLUTION
\footnote{talk presented at the International Symposium on QCD Corrections
and New Physics, Hiroshima, Japan, 1997.}}
\vskip 1.0cm
\centerline{Hsiang-nan Li}
\vskip 1.0cm
\centerline{Department of Physics, National Cheng-Kung University,}\par
\centerline{Tainan, Taiwan, Republic of China}
\vskip 2.0cm
\centerline{\bf abstract}

I review the recent progress in small $x$ physics, concentrating on
the topics relevant to the BFKL evolution.

\vskip 1.0cm
\section{Introduction}

It is known that radiative corrections in perturbative QCD (PQCD) produce
large logarithms, such as $\alpha_s\ln Q$ in the kinematic region with
large momentum transfer $Q$ and $\alpha_s\ln(1/x)$ in the region with
small Bjorken variable $x$. These large logarithms, spoling the
expansion in the coupling constant $\alpha_s$, must be organized in
some way. To sum the various logarithms contained in a parton distribution
function to all orders, four well-known evolution equations have been
proposed: The Dokshitzer-Gribov-Lipatov-Altarelli-Parisi (DGLAP) equation
\cite{AP} sums $\ln Q$ for intermediate $x$, the
Balitskii-Fadin-Kuraev-Lipatov (BFKL) equation \cite{BFKL} sums $\ln(1/x)$
for a small $x$, and the Ciafaloni-Catani-Fiorani-Marchesini (CCFM) equation
\cite{CCFM}, appropriate for both large and small $x$, unifies the above two
equations. In the region with both large $Q$ and small $x$ many gluons
are radiated in scattering processes with small spatial separation among
them, and a new effect from the annihilation of two gluons into one gluon
becomes important. The nonlinear Gribov-Levin-Ryskin (GLR) equation
\cite{GLR} takes into account this effect.
All of the above equations can be applied to QCD processes
at small $x$, though they give different evolutions in $x$ of a parton
distribution function. The DGLAP equation is applicable, since
the relevant splitting functions show the desired $x$ dependence.
In this talk I will concentrate on the BFKL equation.

\section{The BFKL Evolution}

First, I list some important features of the BFKL evolution, which
motivate the studies reported in the following sections.

\subsection{The BFKL Equation}

In the small $x$ region gluon contributions dominate the structure
functions involved in deep inelastic scattering (DIS) of a proton, such as
$F_2(x,Q^2)$, $Q$ being the momentum transfer form the virtual photon.
Since the longitudinal component $\xi p^+$ and the transverse
component $k_T$ of the gluon momentum may be of the same order, the
transverse degrees of freedom of the gluon must be taken into account,
leading to the $k_T$-factorization theorem \cite{J},
\begin{equation}
F_2(x,Q^2)=\int_x^1\frac{d\xi}{\xi}\int \frac{d^2{\bf k}_T}{\pi}
H(x/\xi,Q,k_T)F(\xi,k_T)\;,
\label{ktf}
\end{equation}
where $H$ is the hard photon-gluon scattering subamplitude, and
$F$ the unintegrated gluon distribution function. If $x$ is large,
the $k_T$ dependence in $H$ is negligible. The integration of $F$
over $k_T$ then gives the gluon density $G$,
\begin{equation}
\xi G(\xi,Q^2)=\int_0^Q\frac{d^2 {\bf k}_T}{\pi}F(\xi,k_T)\;,
\end{equation}
and Eq.~(\ref{ktf}) reduces to the conventional collinear factorization
formula.

It is known that leading logarithmic corrections $\alpha_s\ln(1/x)$ are
produced from a ladder diagram, where the rung gluons obey the strong
rapidity ordering,
\begin{equation}
y_1\gg y_2\gg ... \gg y_n\;,
\end{equation}
with $y_1$ the rapidity of the rung gluon closest to the proton, and
$y_n$ the rapidity of the rung gluon closest to the hard scattering.
One further reggeizes the ladder gluons by assuming that their invariant
masses are given by $k^2=-k_T^2$. Summing the Reggeon ladders to all
orders, one obtains the BFKL equation that describes the evolution
of $F$ at small $x$ \cite{BFKL},
\begin{equation}
-x\frac{\partial}{\partial x}f(x,k_T)=
{\bar\alpha}_s k_T^2\int\frac{d^2 {\bf k}'_T}{\pi k_T^{'2}}
\left[\frac{f(x,k'_T)-f(x,k_T)}{|k_T^{'2}-k_T^2|}
+\frac{f(x,k_T)}{(4k_T^{'4}-k_T^4)^{1/2}}\right]\;,
\label{bfkl}
\end{equation}
with $f(x,k_T)=k_T^2F(x,k_T)$ and
${\bar \alpha}_s=N_c\alpha_s/\pi$, $N_c=3$ being the numebr of colors.

\subsection{Solution to the BFKL Equation}

Using the Mellin transform
\begin{eqnarray}
{\bar f}(x,\omega)&=&\int_0^\infty(k^2)^{-\omega-1}f(x,k)dk^2\;,
\\
f(x,k)&=&\frac{1}{2\pi i}\int_{c-i\infty}^{c+i\infty}(k^2)^\omega
{\bar f}(x,\omega)d\omega\;,
\label{imt}
\end{eqnarray}
Eq.~(\ref{bfkl}) in the $\omega$ space is written, for fixed $\alpha_s$, as
\begin{equation}
-x\frac{\partial}{\partial x}{\bar f}(x,\omega)=
K(\omega){\bar f}(x,\omega)\;,
\label{bft}
\end{equation}
where the trasformed BFKL kernel $K$ is given by
\begin{eqnarray}
& &K(\omega)={\bar\alpha}_s[2\psi(1)-\psi(\omega)-\psi(1-\omega)]\;,
\\
& &\psi(\omega)=\frac{d}{d\omega}\ln \Gamma(\omega)\;,\;\;\;\;
\psi(1)=-\gamma_E\;,
\end{eqnarray}
$\gamma_E$ being the Euler constant. $K(\omega)$ has a maximum at
$\omega=1/2$, around which it can be expanded as
\begin{equation}
K\left(\frac{1}{2}+i\nu\right)=
\lambda-\frac{1}{2}\lambda''\nu^2+O(\nu^4)\;,
\label{kap}
\end{equation}
with the constants $\lambda={\bar\alpha}_s4\ln 2$ and
$\lambda''={\bar\alpha}_s28\zeta(3)$, $\zeta(3)=1.202$.
Substituting Eq.~(\ref{kap}) into (\ref{bft}), one solves for ${\bar f}$,
and the inverse transform Eq.~(\ref{imt}) gives 
\begin{equation}
f(x,k)\approx \left(\frac{x}{x_0}\right)^{-\lambda}
\frac{{\bar f}(x_0,1/2)k}{[2\pi\lambda''\ln(x_0/x)]^{1/2}}
\exp\left[-\frac{(\ln k^2-\ln{\bar k}^2)^2}{2\lambda''\ln(x_0/x)}\right]\;,
\label{bfe}
\end{equation}
where
\begin{equation}
\ln{\bar k}^2=i\frac{d}{d\nu}\left[\ln{\bar F}
\left(x_0,\frac{1}{2}+i\nu\right)\right]_{\nu=0}
\end{equation}
is the center of the Gaussian broadening in $\ln k^2$.
$x_0$ is usually chosen as $0.1$, below which the BFKL evolution begins.

Equation (\ref{bfe}) exhibits the following features:

1. It possesses a power-law rise as $x^{-\lambda}$ with
$\lambda\approx 0.5$ for reasonable values of ${\bar\alpha}_s$,
which is attributed to hard pomeron exchanges (a pomeron can
be regarded as a color-singlet Reggeon ladder). Recent HERA data \cite{H1}
of the DIS structure function $F_2$ confirm this feature.

2. It does not depend on $Q$, implying that $F_2$ is insensitive to the
variation of the momentum trasfer. However, the rise of the data
is characterized by $\lambda$ with a stronger $Q$ dependence:
$\lambda\sim 0.2$  for $Q^2\sim 4$ GeV$^2$ and $\lambda\sim 0.5$ for
$Q^2\sim 50$ GeV$^2$. At very low $Q$, $\lambda\sim 0.08$ corresponds to
soft pomeron exchanges.

3. The transverse momentum $k_T$ diffuses into a nonperturbative region
at small $x$. Since the Gaussian width $\sqrt{2\lambda''\ln(x_0/x)}$
increases with $1/x$, the distribution of the gluon in $\ln k^2$ certered
at $\ln{\bar k}^2$ becomes broader. The nonperturbative region with small
$k^2$ then gives essentail contributions to $F_2$, and PQCD is not reliable.

4. The power-law rise of $f$ renders $F_2$ and the DIS cross section
$\sigma$ violate the unitarity bound $\sigma\le {\rm const.}\times
\ln^2(1/x)$.

The above features have stimulated intensive studies of the BFKL
evolution, whose progress will be reported below.

\section{How to explain data?}

To explain the data of the structure function $F_2$, the DGLAP equation
should be combined in some way, which introduces the $Q$ dependence
through the $\ln Q$ summation.

\subsection{The DGLAP Equation}

One may employ the DGLAP equation for the gluon density $G$
directly \cite{AP},
\begin{equation}
Q^2\frac{d}{d Q^2}G(x,Q^2)=\int_x^1\frac{d\xi}{\xi}P_{gg}(x/\xi)
G(\xi,Q^2)\;.
\label{dg}
\end{equation}
The required rise of $G$ at small $x$ is the result of the splitting
function
\begin{equation}
P_{gg}(z)={\bar\alpha}_s\left[\frac{1}{z}+\frac{1}{1-z}-2+z(1-z)\right]\;,
\label{pgg}
\end{equation}
which diverges at $z\to 0$. It can be shown that $G$ evolves according to
\begin{equation}
xG(x,Q^2)\propto \exp\left[2\sqrt{\frac{N_c}{\beta_0}\ln\frac{1}{x}
\ln\frac{t}{t_0}}\right]\;,
\label{dge}
\end{equation}
with the variable $t_{(0)}=\ln(Q_{(0)}^2/\Lambda_{\rm QCD}^2)$ and
$\beta_0$ the first coefficient of the QCD beta function.

For $Q_0^2=4$ GeV$^2$, Eq.~(\ref{dge}) gives $\lambda\sim 0.11$ for
$Q^2\sim 15$ GeV$^2$ and $\lambda\sim 0.15$ for $Q^2\sim 30$ GeV$^2$.
Obviously, these values of $\lambda$ are too small to explain the
rise of $F_2$. Hence, a steeper input $G(x,Q_0^2)$ must be adopted to
compensate the slower DGLAP evolution.  If choosing $Q_0^2=0.3$ GeV$^2$,
$\lambda\sim 0.4$ for $Q^2\sim 20$ GeV$^2$ is large enough, and a flat
iuput serves the purpose \cite{GRV}. However, it has been criticized that
the PQCD evolution around this low $Q_0$ is not reliable \cite{AKMS}.

The $k_T$ factorization shown in Eq.~(\ref{ktf}) is in fact
equivalent to the collinear (mass) factorization \cite{J}, on which the
DGLAP equation is based. Expressing the solution to Eq.~(\ref{bft}) in the
moment space, one obtains
\begin{equation}
{\bar f}_N(\omega)\propto\frac{f(x_0,\omega)}{N-K(\omega)}\;,
\label{bft1}
\end{equation}
with the definition
\begin{equation}
{\bar f}_N(\omega)=\int_0^1dx x^{N-1}{\bar f}(x,\omega)\;.
\end{equation}
Equation (\ref{bft1}) implies a pole of ${\bar f}_N(\omega)$ at
$\omega=\gamma_N$, which satisfies $K(\gamma_N)=N$, and the parametrization
of $F_N(k)$ from Eq.~(\ref{imt}) \cite{J},
\begin{equation}
F_N(k)=\frac{f_N(k)}{k^2}=
\frac{\gamma_N}{\pi k^2}\left(\frac{k^2}{\mu^2}\right)^{\gamma_N}
G_N(\mu^2)\;,
\end{equation}
$\mu$ being an arbitrary factorization scale.
$\gamma_N$ is the BFKL anomalous dimension, whose perturbative expansion
is given by
\begin{equation}
\gamma_N=\frac{{\bar\alpha}_s}{N}+2\zeta(3)
\left(\frac{{\bar\alpha}_s}{N}\right)^4+
O\left(\left(\frac{{\bar\alpha}_s}{N}\right)^6\right)\;.
\end{equation}
Inserting the above expression into Eq.~(\ref{ktf}) in the moment space,
one arrives at
\begin{equation}
F_{2,N}(Q^2)=C_N(Q^2/\mu^2)G_N(\mu^2)\;,
\label{kdg}
\end{equation}
with the coefficient function
\begin{equation}
C_N(Q^2/\mu^2)=\gamma_N\int\frac{d^2{\bf k}}{\pi k^2}
\left(\frac{k^2}{\mu^2}\right)^{\gamma_N}H_N(Q,k)\;.
\end{equation}
It can be shown that $G_N(\mu^2)$ is related to the $N$-th moment of the
gluon density $G$ defined in the modified minimal subtraction scheme
\cite{CH}, and the BFKL anomalous dimension $\gamma_N$ is the same as the
gluon anomalous dimension $\gamma_{gg,N}$, the $N$-th moment of $P_{gg}$
in Eq.~(\ref{pgg}), up to leading logarithms. Obviously, Eq.~(\ref{kdg})
is equivalent to the collinear factorization formula for $F_2$.

\subsection{Next-to-leading Logarithms}

As a more complete analysis, one includes the singlet quark contribution
through the DGLAP equation \cite{EHW}
\begin{equation}
\frac{d}{d\ln \mu^2}\left(\begin{array}{c}
Q\\G
\end{array}\right)=\gamma\left(\begin{array}{c}
Q\\G
\end{array}\right)\;,\;\;\;\;
\gamma=\left(\begin{array}{cc}
\gamma_{qq}& \gamma_{qg}\\
\gamma_{gq}& \gamma_{qg}
\end{array}\right)\;,
\label{ntl}
\end{equation}
where $Q$ denotes the singlet quark distribution function.
The anomalous dimensions $\gamma_{qq}$ and $\gamma_{qg}$, giving the
next-to-leading-logarithm summation, leads to a steeper rise of $G$ at
small $x$ compared to the DGLAP evolution in Eq.~(\ref{dge}), which
takes into account only the leading gluon contribution. The analysis
concludes that a flat input $G(x,Q_0^2)$ at $Q_0^2=4$ GeV$^2$ is preferred.

The above conclusion has been justified in an alternative way.
One extracts the evolution equation for the gluon decaisy $G$ from
Eq.~(\ref{ntl}) \cite{BF},
\begin{equation}
\left[\frac{\partial}{\partial \xi\partial\zeta}+
\delta\frac{\partial}{\partial \xi}-\gamma^2\right]G(\xi,\zeta)=0\;
\label{ds}
\end{equation}
with
\begin{eqnarray}
& &\xi=\ln\left(\frac{x_0}{x}\right)\;,\;\;\;\;
\zeta=\ln\left(\frac{t}{t_0}\right)\;,
\nonumber\\
& &\delta=\frac{11+2n_f/27}{\beta_0}\;,\;\;\;\;
\gamma=\sqrt{\frac{12}{\beta_0}}\;,
\end{eqnarray}
where the variable $t$ has appeared in Eq.~(\ref{dge}). With the further
variable changes
\begin{equation}
\sigma=\sqrt{\xi\zeta}\;,\;\;\;\;
\rho=\sqrt{\frac{\xi}{\zeta}}\;,
\end{equation}
Eq.~(\ref{ds}) is easily solved to give
\begin{equation}
G(\sigma,\rho)\propto \frac{1}{\sqrt{4\pi\gamma\sigma}}
\exp\left(2\gamma\sigma-\delta\frac{\sigma}{\rho}\right)\;.
\label{g}
\end{equation}
Equation (\ref{g}) exhibits two asymptotic scaling laws:
\begin{eqnarray}
& &\ln G\propto \sigma \hspace{1.5cm} {\rm for}\;\;{\rm fixed}\;\; \rho\;,
\\
& &\ln G\propto {\rm const.}\hspace{1.0cm} {\rm for}
\;\;{\rm fixed}\;\; \sigma\;.
\end{eqnarray}
The plots of the data of $\ln F_2$ with respect to $\sigma$ and $\rho$
indeed confirm the double scaling bebavior at large $\sigma$ and
$\rho$. On the other hand, different inputs of $G(x,Q_0^2)$ lead to
different slopes of the scaling. It was found that the predictions from
the flat input match the data better. Because of the above more complete
analyses, one is tempted to conclude that soft pomeron exchanges dominate
in the low $Q$ region. It was also argued that the BFKL pomeron appears
only for \cite{BF}
\begin{equation}
\ln\left(\frac{1}{x}\right) > \left[\frac{\alpha_s(Q_0)}
{\alpha_s(Q)}\right]^{20}\;.
\end{equation}

\subsection{The CCFM Equation}

At last, one may resort to the CCFM equation, which embodies both the
$\ln Q$ and $\ln(1/x)$ summations. It is written as \cite{CCFM}
\begin{eqnarray}
F(x,p_T,Q)&=&F^{(0)}(x,p_T,Q)+\int_x^1 dz
\int\frac{d^2q}{\pi q^2}\theta(Q-zq)
\Delta_S(Q,zq)
\nonumber \\ 
& &\hspace{2.0cm}\times
{\tilde P}(z,q,p_T)F(x/z,|{\bf p}_T+(1-z){\bf q}|,q)\;,
\label{ccfm}
\end{eqnarray}
with the function 
\begin{equation}
{\tilde P}={\bar\alpha_s}(p_T)
\left[\frac{1}{1-z}+\Delta_{NS}(z,q,p_T)\frac{1}{z}+z(1-z)\right]
\label{pgg1}
\end{equation}
similar to the gluon splitting function $P_{gg}$ in Eq.~(\ref{pgg}).
The so-called ``Sudakov" exponential $\Delta_S$ and the ``non-Sudakov"
exponential $\Delta_{NS}$ are given by
\begin{eqnarray}
\Delta_S(Q,zq)&=&\exp\left[-{\bar\alpha_s}
\int_{(zq)^2}^{Q^2}\frac{dp^2}{p^2}
\int_{0}^{1-p_T/p}\frac{dz'}{1-z'}\right]
\nonumber \\
\Delta_{NS}(z,q,p_T)&=&
\exp\left[-{\bar\alpha_s}\int_{z}^{z_0}\frac{dz'}{z'}
\int_{(z'q)^2}^{p_T^2}\frac{dp^2}{p^2}\right]\;.
\label{nons}
\end{eqnarray}
where the upper bound $z_0$ of the variable $z'$ takes the
values \cite{CCFM,KMS},
\begin{eqnarray}
   & & 1 \hspace{1.5cm}  {\rm if}\;\; 1\le (p_T/q)
\nonumber \\
z_0&=& p_T/q \hspace{1.0cm}  {\rm if}\;\; z < (p_T/q) < 1
\nonumber \\
   & & z \hspace{1.5cm}   {\rm if}\;\; (p_T/q)\le z\;.
\end{eqnarray}   
$\Delta_S$ collects the contributions from the ladder diagrams with rung
gluons obeying the strong angular ordering, and is the result of
the $\ln Q$ summation. Those gluons which do not obey the angular
ordering are grouped into $\Delta_{NS}$. In the small $x$ region one
adopts the approximate version of the CCFM equation \cite{KMS},
\begin{eqnarray}
F(x,p_T,Q)&=&F^{(0)}(x,p_T,Q)+{\bar \alpha}_s(p_T)\int_x^1 \frac{dz}{z}
\int\frac{d^2q}{\pi q^2}\theta(Q-zq)\theta(q-\mu)
\nonumber \\ 
& &\hspace{1.5cm}\times
\Delta_{NS}(z,q,p_T)
F(x/z,|{\bf p}_T+(1-z){\bf q}|,q)\;.
\label{cca}
\end{eqnarray}
where the extra function $\theta(q-\mu)$ introduces an infrared cutoff of
the variable $q$, and $F^{(0)}$ is a flat nonperturbative driving term.

The CCFM evolution with $x$ has been extracted from Eq.~(\ref{cca})
\cite{LL}, given by
\begin{eqnarray}
F\propto \exp[2{\bar \alpha}_s\ln(Q/\mu)\ln(1/x)]\;.
\end{eqnarray}
It shows a steeper rise at small $x$ than the DGLAP evolution in
Eq.~(\ref{dge}), though it also considers only the leading gluon
contribution, and a more $Q$-dependent rise than the BFKL evolution in
Eq.~(\ref{bfe}). This is the reason the CCFM equation can explain the data
well \cite{KMS}. The steeper CCFM rise compared to the DGLAP rise is
attributed to the different choices of the argument of the running coupling
constant in the splitting function, which is $p_T$ in the former
(see Eq.~(\ref{pgg1})) and $Q$ in the latter. Therefore, $\alpha_s(p_T)$
does not run in fact, as the variable $q$ is integrated over in the CCFM
equation (\ref{cca}). While $\alpha_s(Q)$ runs, when solving the DGLAP
equation. Because of $\alpha_s(p_T) > \alpha_s(Q)$, the CCFM evolution
is stronger.

\section{Where is the hard pomeron?}

In the previous section it has been stated that soft pomeron contributions
dominate in the low-$Q$ region, and the high-$Q$ behavior of the DIS
structure function is the consequence of the DGLAP evolution plus a
nonperturbative input. Though the BFKL equation gives an equally good
explanation of the data at large $Q$, it is still worthwhile to
ask where the clear evidence for the BFKL (hard) pomeron is. To answer this
question, many proceses have been proposed.

\subsection{Two-jet Processes}

Consider the two-jet production from hadron-hadron collisions \cite{MN},
\begin{equation}
p+p\to 2 \;\;{\rm jets} +X\;.
\end{equation}
If tagging the jet with longitudinal momentum fraction $x_i$, transverse
momentum $p_{iT}$ and rapidity $y_i$, $i=1$, 2, the momentum fractions
of the initial-state partons will be fixed at $x_i$ due to the kinematic
relation
\begin{equation}
x_i=(p_{iT}/\sqrt{s})\exp(y_i)\;,
\end{equation}
$s$ being the center-of-mass
energy, when the rapidity gap $\Delta y=y_1-y_2$ is large. It was argued
that many minijets are radiated from the $t$-channel exchanged gluon
between the initial-state partons at large $\Delta y$, such that the
cross section $d^2\sigma/(dy_1dy_2)$ increases with 
$\Delta y$. This rise is definitely attributed to the BFKL pomeron
contribution, since the nonperturbative parton distribution functions
take the values at $x=x_i$ (no evolution in $x$) after the jet tagging.
However, an explicit evaluation of $d^2\sigma/(dy_1dy_2)$ showed that
it, contrary to the expectation, decreases with $\Delta y$ in the available
range of $s$ \cite{DDS}. The reason is that $x_i$ approaches unity for
finite $p_{iT}$ and $s$ as $\Delta y$ increases, and thus the parton
distribution functions vanish.

An alternative was then suggested, where the angle decorrelation between
the two tagged jets is measured \cite{DDS}. The radiation of the minijets
washes out the correlation of the two jets with angle $\phi=\pi$ at a
rate that increases with $\Delta y$. The explicit analyses of the $\phi$
decorrelation have confirmed this expectation \cite{DDS}.

\subsection{One Jet in DIS}

A similar process with one forward jet in DIS was proposed \cite{M}.
By tagging the jet with momentum fraction $x_j$ and transverse momentum
$k_{jT}$, the momentum fraction of the initial-state parton is fixed
at $x_j$. $k_{jT}$ is chosen as being large enough to avoid the
$k_T$ diffusion into the nonperturbative region. The $t$-channel exchanged
gluon radiated from the parton emitts many minijets, such that the
structure function $F_2$ behaves as
\begin{equation}
\frac{\partial F_2(x)}{\partial \ln(1/x_j)\partial k_{jT}}
\propto \alpha_s(k_{jT})x_j f(x_j)\left(\frac{x}{x_j}\right)^{-\lambda}\;.
\end{equation}
Note that the parton distribution function
\begin{equation}
f=G+\frac{4}{9}(Q+{\bar Q})
\end{equation}
is evaluated at $x=x_j$, and the nonperturbative ambiguity is removed. The
rise $x^{-\lambda}$ is then clearly identified as the consequence of the
BFKL evolution. Though it is challenging to tag a forward jet,
experimental studies have been attempted with encouraging results
\cite{HA}.

Instead of tagging a forward jet, it has been suggested to tag a forward
photon \cite{KLM}. The advantage of this process is that
it is cleaner. However, it is suppressed by the coupling constant
$\alpha_{EM}$ compared to the previous process, and the photon needs to be
isolated from $\pi^0$.

\subsection{Levin's Viewpoint}

Though all the above analyses and experiments have not yet fully confirmed
the existence of the BFKL pomeron, Levin argued that its evidence had been
contained in the current data \cite{L}. To demonstrate his viewpoint, an
average of the BFKL anomalous dimension $\langle\gamma\rangle$ and the
strength of the shadowing corrections $\kappa$ are defined by
\begin{equation}
\langle\gamma\rangle=\frac{\sum_N\gamma_N G_N(Q^2)}
{\sum_N G_N(Q^2)}\;,\;\;\;\;
\kappa=\frac{3\pi\alpha_s}{Q^2 R^2}G(x,Q^2)\;.
\end{equation}
The BFKL anomalous dimension $\gamma_N$ and the moment of the gluon
density $G_N$ that obeys the DGLAP evolution have been introduced in
Sect. 3.1. $\langle\gamma\rangle$ can be regarded as measuring
the BFKL content in the DGLAP gluon density employed in the explanation
of the DIS data \cite{FRT}. If $\langle\gamma\rangle$ is located between
1/2 and 1, it is claimed that the gluon density contains significant BFKL
pomeron contribution. From the GLR equation (see Eq.~(\ref{glr}) below),
it is easy to observe that $\kappa$ corresponds to the nonlinear term,
describing the effect of two gluon annihilation into one gluon. $R$ is the
correlation length of the radiated gluons. For $\kappa>1$, the shadowing
correction is considered to be large.

A simple analysis showed that the BFKL region between
$\langle\gamma\rangle=1/2$ and $\langle\gamma\rangle=1$ in the
$\ln(1/x)$-$\ln Q^2$ plane indeed penetrates the region which has been
explored by the HERA experiments. However, the BFKL region is completely
located in the region with $\kappa>1$.  Hence, Levin concluded
that the BFKL pomeron contribution has been contained in the current HERA
data, but is suppressed by the shadowing correction, and thus
unobservable.

\section{How to Recover Unitarity?}

As stated in Sect. 2, the power-law rise $x^{-\lambda}$ violates the
unitarity bound, implying that the BFKL equation derived from
the leading-twist and leading-logarithm approximation needs to be
corrected. The attempts to include higher-twist and next-to-leading
logarithmic contributions have been made.

\subsection{The GLR Equation}

In the region with both large $Q$ and small $x$, many gluons are
radiated by partons with small spatial separation among them. A new effect 
from the annihilation of two gluons into one gluon is then essential. 
Taking into account this effect, the BFKL equation is modified by a 
nonlinear term, leading to the GLR equation \cite{GLR}, 
\begin{equation}
\frac{\partial^2xG(x,Q^2)}{\partial\ln(1/x)\partial \ln Q^2}
={\bar\alpha_s}xG(x,Q^2)-\frac{\gamma\alpha_s}{Q^2R^2}
[xG(x,Q^2)]^2\;,
\label{glr}
\end{equation}
The constant $\gamma=81/16$ is regarded as the effective coupling of the 
annihilation process, and the radius $R$ characterizes the correlation 
length of the radiative gluons as mentioned before. It is
obvious that the second term is of higher twist, but becomes
important as $R$ is small. The minus sign in front of it implies that the
annihilation decreases the number of gluons, and that the rise of the gluon
density at small $x$ due to the first term might saturate.

This equation is, however, nonlinear, and the values of $R$ depend on
models. It is set to 2 GeV$^{-1}$ for the model with the radiative gluons
concentrated at hot spots, and 5 GeV$^{-1}$ for the model with
the gluons uniformly distributed in a hadron.

\subsection{Next-to-leading Logarithms}

To recover the unitarity in the framework of the BFKL equation, it
was proposed to compute the kernel that includes next-to-leading
$\ln(1/x)$, which arise from relaxing the strong rapidity ordering of
rung gluons. That is, the contribution from the region with, for example,
\begin{equation}
y_1\sim y_2 \gg y_3 \gg ... \gg y_n\;,
\end{equation}
should be computed. The virtual and real corrections have been obtained in
\cite{F} and in \cite{VDD}, respectively. However, an explicit expression
of the BFKL kernel including next-to-leading logarithms has not been
extracted from the above analyses (the phase-space integrals for the
amplitudes of real gluon emissions given in \cite{VDD} were not performed).
Hence, no concrete conclusion on the unitarity has been drawn. Since the
involved calculations are very tedious, I will not go into the details.
Readers are referred to Ref.~\cite{CC} for summary of the progress on this
subject.

\subsection{Multiple Pomeron Exchanges}

Multiple pomeron exchanges, as higher-twist contributions, have been
studied by means of colored-dipole scatterings \cite{MP}. Consider an
onium, which is the $Q{\bar Q}$ bound state with $Q$ a heavy quark. The
gluon emission inside the onium, from the viewpoint of color flow, can be
regarded as being composed of the $Q$-${\bar Q}$ pair in the color octet
state. Many gluons then imply many quark pairs. The emitted quarks
and valence quarks, combined among each other, form many colored-dipoles
in the oniun. In this picture, the onium-onium forward scattering is
then formulated as multiple dipole scatterings. One of the advantages
of this process is that the onium size provides a natural infrared cutoff,
and thus resolves the problem of the $k_T$ diffusion.

To compute the cross section, one defines a dipole density $n(x,Y)$ for an
onium, with $x$ the transverse size of a dipole and $Y$ the rapidity. At
high energy, it has been shown that $n$ increases with $Y$ due to the same
mechanism as of the rise of the gluon distribution function with $1/x$.
That is, the equation that governs the evolution of $n$ with $Y$ is
equivalent to the equation that governs the evolution of $F$ with
$\ln(1/x)$ \cite{CM}. However, the derivation of the kernel for the former
is much simpler than that for the latter. Though the whole evolution
kernels for $n$ and for $F$ are equal, the parts from real gluon emissions
and from virtual gluon emissions differ. To obtain the kernel for $F$, all
the higher-order diagrams are considered. For the kernel of $n$, only the
diagrams with radiative gluons emitted before the hard onium-onium
scattering are considered, since they are the diagrams that modify the
dipole density for an initial-state onium. With this criterion, the number
of diagrams needed to be computed is greatly reduced. This is another
advantage of the above process.

Since the formalism for dipole scatterings is simpler, it can be extended
to include multiple pomeron exchanges. Recall that the $t$-channel gluon
responsible for the hard scattering is reggeized at large rapidity
gap, and two reggeons form a pomeron.  The cross sections including
single and double pomeron exchanges have been obtained, which exhibit
a rise with $Y$ \cite{CM}. As to the unitarity, for which infinite many
pomeron exchanges may be essential, no conclusive progress has
been made following this vein.

\subsection{Reggeon Compound States}

The higher-twist contributions from multiple pomeron (reggeon) exchanges
can be taken into account in another interesting approach \cite{K}. As
explained in Sect. 2, a reggeized gluon is a two-dimensional object, since
its propagator depends only on the transverse momentum. Hence, in
high-energy scatterings (3+1) QCD is reduced to (2+1) Regge theory
\cite{LN}. Consider the compound state of $N$ reggeons distributed randomly
in a $x$-$y$ plane. It has been shown that only the nearest-neighbor
interaction survives in the large $N_c$ limit. Under this approximation,
these reggeons can be arrayed into a one-dimensional space in the sequence
where the interaction occurs between any two neighboring reggeons.
Certainly, there exist two such one-dimensional spaces, characterized by
$z=x+iy$ and ${\bar z}=x-iy$. Then the (2+1) Regge theory is further
reduced to a $(1+1)\otimes (1+1)$ Schrodinger equation. It is trivial to
observe that each of the one-dimensional systems is an exactly
solvable Heisenberg model with spin $s=0$.

The Hamiltonian and the corresponding Schrodinger equation are
written as
\begin{eqnarray}
& &H_N(z,{\bar z})={\bar\alpha}_s\left[H_N(z)+H_N({\bar z})\right]\;,
\\
& &H_N\chi_N=E_{N}\chi_N\;,
\end{eqnarray}
where $\chi_N$ is the wave function of the $N$-reggeon compound state with
the eigenenergy $E_{N}$. The eigenenergy can be identified as the
intercept of the Regge trajectory. Therefore, by computing the eigenenergy
of the compound state, one understands how the power-law rise of the
structure function $F_2$ varies with the reggeon number $N$:
\begin{equation}
F_2^{(N)}(x,Q^2)\sim x^{-E_N}\;.
\end{equation}
$E_2={\bar\alpha}_s 4\ln 2$ for the $N=2$ case, which corrsponds
to the BFKL pomeron exchange, was found. The tendency $E_2 > E_3 > ...$
has been observed, indicating that the rise is indeed softened gradually as
$N$ increases, though it is still power-like. Since the $N\to \infty$ case
remains unsolved, a concrete conclusion on the unitarity is still not
available.

\section{Summary}

In this talk I reviewed the recent progress in small $x$ physics.
Obviously, many issues need to be studied theoretically and experimentally
with more effort.

\section*{Acknowledgment}
This work was supported by the National Science Council of Republic
of China under the Grant No. NSC87-2112-M-006-018.

\end{document}